\documentclass[twocolumn]{autart}
\usepackage{epsfig}
\usepackage{amsfonts,amssymb,amsmath}
\usepackage[comma,numbers,square,sort&compress]{natbib}
\usepackage{graphicx}
\usepackage{bbm}
\usepackage{stmaryrd}
\usepackage{colortbl}

\usepackage{graphicx,amsmath,amssymb,remark,color}
\usepackage{bm}
\usepackage{algorithm} 
\usepackage{algorithmic}
\usepackage{graphicx}
\usepackage[outdir=./]{epstopdf}
\graphicspath{{../eps/}}
\DeclareGraphicsExtensions{.eps}

\newenvironment{Proof}{\noindent{\em Proof:\/}}{\hfill $\Box$\par}
\newtheorem{Example}{Example}
\newtheorem{Remark}{Remark}[section]

\newtheorem{Definition}{Definition}[section]
\newtheorem{Theorem}{Theorem}[section]

\newtheorem{Lemma}{Lemma}[section]
\newtheorem{Assumption}{Assumption}[section]

\DeclareMathOperator{\modd}{mod}
\definecolor{ao}{rgb}{0.13, 0.55, 0.13}

\newcommand{\bSigma}{{\mathbf \Sigma}}

\newcommand{\q}{{\rm q}}
\newcommand{\sgn}{{\rm sgn}}

\newcommand{\ba}{{\bf a}}

\newcommand{\by}{{\bf y}}
\newcommand{\bv}{{\bf v}}
\newcommand{\cE}{{\mathcal E}}

\newcommand{\cD}{{\mathcal D}}

\newcommand{\cI}{{\mathcal I}}

\begin{document}

\begin{frontmatter}

\title{Control of Large-Scale Networked Cyberphysical Systems Using Cryptographic Techniques}

\author{Yamin Yan, Zhiyong Chen, and Vijay Varadharajan}
\address{School of Electrical Engineering and Computing, 
The University of Newcastle\\
Callaghan, NSW 2308, Australia, 
Tel: +61 2 4921 6352, Fax: +61 2 4921 6993\\
Emails:  {\tt yamin.yan, zhiyong.chen, vijay.varadharajan@newcastle.edu.au}}

 \begin{abstract}
This paper aims to create a secure environment for networked control systems composed of multiple dynamic entities and computational control units via networking, in the presence of disclosure attacks.  In particular, we consider the situation where some dynamic entities or control units are vulnerable to attacks and can become malicious. Our objective is to ensure that the input and output data of the benign entities are protected from the malicious entities as well as protected when they are transferred over the networks in a distributed environment. Both these security requirements are achieved using cryptographic techniques. However, the use of cryptographic mechanisms brings additional challenges to the design of controllers in the encrypted state space; the closed-loop system gains and states are required to match the specified  cryptographic algorithms.  In this paper, we propose a methodology for the design of secure networked control systems integrating the cryptographic mechanisms with the control algorithms. The approach is based on the  separation principle, with the cryptographic techniques addressing the security requirements and the control algorithms satisfying their performance requirements.  
\end{abstract}

\begin{keyword}
Networked control systems,  Large-scale systems, Cyberphysical security, Encryption, Quantization,
Stabilization \end{keyword}

\end{frontmatter}

\section{Introduction}

Large-scale systems (LSSs) have been rapidly developed in interdisciplinary fields with the 
technological advances in computer science, software engineering, and systems engineering. 
The concept of LSS adopted in this paper involves dynamic
systems involving a large number of states, control inputs, and parameters; see, e.g.,  \cite{mehraeen2011,baldi2014}.
Such systems usually consist of a class of  interconnected subsystems with physical coupling among them. 
It also includes the class of multi-agent systems and complex systems. In these systems, 
people are more interested in their collective phenomena  and behaviors.
A common feature of these LSSs is that they have multiple  dynamic entities. 
Therefore, attacks can occur on a partial set of entities rather than the whole system, which necessitates the need to protect parts of the system even though some other parts are under attack.

%
%

For either a small-scale or a large-scale system, the controller may be implemented on a remote computing unit
if the system itself does not have local computing facilities. Such an architecture is known as a networked control system (NCS) \cite{nesic2009unified,walsh2002stability,sandberg2015cyberphysical}. It is also called a cyberphysical system especially as it involves physical dynamics
and computer based algorithms via networking.  
The main challenges in studying NCSs lie in 
understanding the control principles with sampling, quantization, communication time-delay, packet dropout, etc. 
NCSs can find extensive applications in different fields, including smart grids, transportation, smart cities, and so on.


The present research is on large-scale systems with networked control (LSS-NC), which is
the synthesis of the aforementioned two concepts and involves all the associated research issues. 
It has become a class of representative complicated engineering systems involving various state-of-the-art technologies. 
For instance, controllers for a group of drones lacking  onboard computing capacity are usually 
implemented on decentralized remote computing base stations \cite{otto2018optimization, cummings2015operator}.  A railway system is a typical large-scale system
with strong physical coupling among trains, whose dispatch planning and control is typically implemented in remote central stations
through various network communication \cite{kiessling2018contact}.
The main target of this research is not on the traditional control issues for LSSs and/or NCSs but 
 on the need to develop a secure environment. LSS-NC is indeed a niche platform for the research 
from the security perspective because it allows us the possibility of protecting a whole system even when
its partial dynamic entities or remote control units are subject to attack.

Security and privacy have become critical issues for modern industrial systems.
Improperly addressing these issues could result in massive economic losses or even  threats of human safety.
Specifically,  attack may occur ubiquitously and we consider the circumstance for LSS-NC
that some dynamic entities
or control units are hacked and can become malicious. The secure environment 
studied in this paper involves the system entities having mechanisms to protect their private input/output information from the malicious entities
and controllers. 

 
In the literature, the significance of addressing the cyberphysical security of NCSs has been widely recognized \cite{cardenas2008research,smith2015covert,amin2013security,teixeira15,mo2015physical,dibaji2019systems}.  
For example, cyberattacks such as Stuxnet malware \cite{falliere2011} 
to typical NCSs operated through 
supervisory control and data acquisition (SCADA)  systems have been well studied in, e.g., \cite{gian09, gorman2009}.
Researchers are interested in knowing how much damage attacks can cause by 
injecting malicious signals or commands into an NCS and what can be done to mitigate that damage \cite{cardenas2011,zhu2014performance,teixeira2015}. 
{  The research in this paper is mainly concerned about data confidentiality violation by disclosure attacks.} 
There has been an increasing interest for development of 
sophisticated control and estimation algorithms with encrypted data for defending disclosure attacks
in \cite{kogiso2015cyber,farokhi2016secure,kim2016encrypting,kim2019encrypted,MZamani} among others.  
 


Also, the research on secure LSSs has been attempted in recent literature, in particular, on consensus of MASs \cite{WFang, duan2015privacy, ruan2019secure, mo2017privacy}.  Privacy and security challenges associated with consensus protocol
are of great practical and theoretical importance in a wide range of applications. An MAS requires
information exchange via network to achieve consensus. Meanwhile, it is usually important to
avoid information leakage during information exchange in a secure control environment.
Apart from these cryptography based approaches,  noise-based obfuscation approaches
are also well studied that mask the true state values by adding or subtracting random noises to the control process. The research along this line has also been rapidly developed over the past years; see, e.g., \cite{kefayati2007secure,nozari_ifac_2015,mo2017privacy}. 


 In the design of secure networked control systems, we consider two specific security requirements. The first requirement is concerned with the protection of information being transferred over the networks. For instance, the input and output data transferred between the system (plant) and the controller over untrusted networks. The second requirement is concerned with the protection of data between the components within a system. For instance, consider a system consisting of say three subcomponents A, B and C. This requirement is concerned with securing the data generated by the component A from the components B and C. Similarly for data generated by components B and C. Such protection will be needed if there is mutual distrust between the various components within a system. Hence the first requirement addresses protection between systems, whereas the second requirement is concerned with protection within a system. To achieve both these security requirements, we have developed a double-layer security scheme using cryptographic mechanisms. One cryptographic layer offers protection against attacks on external networks, connecting the system plant(s) with the control units. The other layer provides cryptographic mechanisms offering protection of data from each of the components within the system.

Compared to the aforementioned literature on secure control of NCSs, our contribution proposes an approach to security that is more comprehensive. Often the works on NCSs have not differentiated the system entities, only considering the case when the remote controller is attacked, which is the first security objective. On the other hand, the works on the LSSs have mainly focused on the controller being synthesized within the entity, thereby not addressing separately attacks on the controller, which is only the second security objective.

When it comes to the implementation of the two security layers, in general, one has a choice of several cryptosystems to choose from.  
In this paper, we use the RSA \cite{rivest1978method} and Paillier \cite{paillier1999public} cryptosystems to realize the two security layers, as both these systems are well known and have been used in many research works in control systems. In terms of formal security properties, though Paillier is semantically secure, algorithms such as RSA are not. The notion of semantic security implies that any information revealed cannot be feasibly extracted and is equivalent to the property of ciphertext indistinguishability under chosen-plaintext attack. However, algorithms such as RSA can be made to be semantically secure through the use of random encryption padding schemes such as Optimal Asymmetric Encryption Padding~\cite{extra}. For the purposes of this paper, it is sufficient to argue that the schemes we have used are informally secure in that given a ciphertext and the public encryption key, a computationally bounded adversary will not be able to determine the actual plaintext or at least the probability of this happening can be made to be very low using well-constructed public key systems (e.g. for RSA, choosing suitable design parameters such as appropriate large primes and suitable exponents etc.).


The contributions of this paper are as follows.
First, we propose a novel double-layer security scheme using cryptographic mechanisms to satisfy both of our security objectives mentioned above. We instantiate such a double-layer scheme by Paillier-RSA encryption techniques. 
The original signals coming from different components in a system are encrypted in an inner layer
to protect the transfer of data over an insecure communication network. Then there is an outer-layer protecting the data between the components. 
{  The inner-layer offers protection against external attackers 
and the outer-layer offers protection between various system components.
The combination of both these protection layers provides security of the overall networked control system.}


Second, a systematic methodology is proposed for the design of a secure networked control system based on the principle of separation consisting of four procedures, namely, quantization, encryption, control, and decryption.
This is being decomposed into two independent sub-problems, namely, a nonencrypted control problem and a quantization-encryption-decryption problem. Both sub-problems should be characterized in the same bounded space as described in Section~\ref{framework} to avoid computation overflow or underflow. We provide a rigorous mathematical proof illustrating this. Another characteristic of this cryptographic design strategy is that it is independent of control problem. That is, we can separately handle different controller design problems to achieve different desired control objectives while simultaneously achieving the security objectives.



Third, we apply the proposed double-layer cryptographic mechanisms and the secure control approach to solve a class of practical problem, namely, the encrypted centralized control problem for stabilization of multi-input multi-output (MIMO) systems.
The effectiveness of the proposed design is illustrated theoretically and by numerical simulation.

The remainder of the paper is organized as follows. Section~\ref{pre} presents preliminary concepts and algorithms. Section~\ref{formulation} gives the rigorous problem formulation. Section~\ref{framework} presents a general 
encrypted LSS-NC architecture.
A specific controller design for the stabilization problem of MIMO systems is given in Section~\ref{case1} 
followed by a numerical example.  Finally, we close the paper with some concluding remarks in Section~\ref{conclusion}. 

\medskip

{\it Notation.}  The vector lumped by the vectors $c_1, \cdots, c_r$ is denoted by
$\mbox{col} (c_1, \dots, c_r) = [c_1^T, \dots, c^T_r]^T$. 
For a set $\mathcal S$, its element $a$ is denoted by $a\in \mathcal S$.
For a vector $a=[a_1,\cdots, a_r]^T$, writing $a \inplus \mathcal S$ means that $a_i \in  \mathcal S, i=1,\cdots, r$.
For a scalar variable and scalar valued function $\cE$, element-wise operation is used throughout the paper, that is,
$\cE(a) = [\cE(a_1),\cdots, \cE(a_r)]^T$.  For a vector $a$, its Euclidean norm is denoted by $\|a\|$ and
its maximum norm by  $\|a\|_\infty$.  For a real matrix $A$, its  induced 2-norm is denoted by $\|A\|$, and its element-wise
absolute value by $|A|$. Let $\mathbf{1}$ be a column vector whose elements are all $1$ and $I$ an identity matrix 
and dimension is from the context.

\section{Preliminaries}\label{pre}

Using cryptographic algorithms to encrypt and decrypt messages during the control process provides the technical foundation 
for creating a secure  environment.  
In this section, we first revisit some preliminaries of RSA~\cite{rivest1978method} and Paillier~\cite{paillier1999public}. 
Quantization of signals, as pre-requisite for  signal encryption,  is also discussed. 

 RSA is a well-known public-key cryptosystem, based on the computational difficulty of factorizing a large composite number.
In RSA, the encryption key is public and it is different from the decryption key, which is kept secret (private). 
The encryption process can be denoted by a function $\cE_R: \mathcal{Z}_{n_R} \mapsto \mathcal{Z}_{n_R}$ for a prescribed positive integer $n_R$ and 
the decryption process by its inverse $\cE_R^{-1}$. Throughout the paper, we 
denote the set of integers modulo $n$ as  $\mathcal{Z}_{n} =[0, 1,\cdots, n-1]$.


%
%
%
%
%
%
%

The Paillier cryptosystem, named after the inventor Pascal Paillier,  is a probabilistic asymmetric algorithm for public key cryptography. It is based on the fact that computing $n$-th residue classes is computationally difficult. 
Also, in Paillier cryptosystem, the encryption key is public and it is different from the decryption key which is kept secret (private). 
The encryption process can be denoted by a function $\cE_P: \mathcal{Z}_{n_P} \mapsto \mathcal{Z}^*_{n_P^2}$ for a prescribed positive integer $n_P$ and  the decryption process by its inverse $\cE_P^{-1}$. Here, $\mathcal{Z}^*_{n} =\{a\in \mathcal{Z}_{n}\; |\; \mbox{gcd}(a,n)=1 \}$. The Paillier cryptosystem has 
an additive homomorphic property $\cE_p(m_1+m_2) =\cE_p(m_1) \cE_p(m_2) \modd n_P^2$ for $m_1, m_2, m_1+m_2 \in \mathcal{Z}_{n_P}$. 
This property means that, given only the public key and the encryption of $m_1$ and $m_2$, one can compute the encryption of $m_1+m_2$.


%
%
%
%
%
%

The computation associated with encryption and decryption procedures in the aforementioned cryptosystems is performed on 
integers. To facilitate the procedures in control systems, we need to apply a function that converts an analogue signal into 
integers. For this purpose, we define a set of signed fixed-point rational numbers in base~$2$, $\mathcal{Q}_{(n,m)}$, that 
 contains all  rational numbers in  $[-2^{n-m-1},2^{n-m-1}-2^{-m}]$ separated from each other by $2^{-m}$, that is, 
\begin{align*}
		\mathcal{Q}_{(n,m)}:=\Big\{-b_n2^{n-m-1}+\sum_{i=1}^{n-1}2^{i-m-1}b_i\; \\ | \; b_i\in\{0,1\}, \forall i\in\{1,\dots,n\}\Big\}.
\end{align*}
The positive integers $n,m$ determine the length and the resolution of the fixed-point rationals. Moreover, a mapping 
 $\cI_{n,m}:\mathcal{Q}_{(n,m)}\mapsto \mathcal{Z}_{2^n}$ is defined to convert the rationals into integers, with 
  $\cI_{n,m}(a)=2^m a \modd 2^n, \forall a\in\mathcal{Q}_{(n,m)}$.  This mapping is bijective with 
  the  inverse mapping $\cI_{n,m}^{-1}:\mathcal{Z}_{2^n}\mapsto \mathcal{Q}_{(n,m)}$ expressed as 
 \begin{equation*}
  \cI_{n,m}^{-1}(a)= \left\{ \begin{array}{ll}
  (a-2^{n})/2^m, & a\geq 2^{n-1} \\
  a /2^m, & {\rm otherwise}   
  \end{array}\right. ,\forall a\in\mathcal{Z}_{2^n}.
  \end{equation*}

In summary,  there are four elementary functions used in the problem formulation, representing rational-integer conversion
and encryption, as follows:
\begin{align}
\cI_{n,m} & :\mathcal{Q}_{(n,m)}\mapsto \mathcal{Z}_{2^n} \nonumber \\
\cE_P &:  \mathcal{Z}_{n_P} \mapsto  \mathcal{Z}_{n_P^2}, \;\; n_P > 2^n \nonumber \\
\cE_{R,i} & :  \mathcal{Z}_{n_R} \mapsto  \mathcal{Z}_{n_R}  \nonumber  \\
\bar \cE_{R,i} &: \mathcal{Z}_{n_R} \mapsto \mathcal{Z}_{n_R},\;n_R \geq n_P^2 ,\; i=1,\cdots, N \label{encrypt}
  \end{align}
 where the subscripts $P$ and $R$ represent Paillier and RSA, respectively, as further explained 
in Theorem~\ref{thm-crypto}. 
The specific selections of these functions, including the encryption keys, will be elaborated in the next section. 
Based on these functions, we define the following composite functions, for the convenience of presentation:
    \begin{align}
 \cE_o(s) &=   \cE_{P} (\cI_{n,m}(s) )\nonumber \\
  \cE_i(s) &= \cE_{R,i}( \cE_o(s))\nonumber \\
\cD_i(s)  &= \cE_{R,i}^{-1}(s) \nonumber \\
\bar \cE_{i} (s) &=\bar \cE_{R,i}(s) \nonumber\\
\cD_o(s) &= \cI^{-1}_{n,m}(\cE_{P}^{-1} (s) \modd 2^n ) \nonumber\\
 \bar \cD_i (s) &=\cD_{o} (\bar \cE_{i}^{-1}(s)).
  \end{align}


\section{Problem Formulation}\label{formulation}
We consider a large-scale control system composed of $N\geq 1$ {  entities (components)} whose dynamics are given as follows 
\begin{align}
x_i (k+1) &= f_i (x(k), u_i(k))  \nonumber \\
y_i(k) &= h_i(x_i(k), u_i(k)), \; i=1,\cdots,N, \label{plant}
\end{align}
where $x_i(k) \in {\mathbb R}^s$ is the state, $u_i(k)\in{\mathbb R}^p$ the input, and $y_i(k)\in {\mathbb R}^q$ the output, of the $i$-th entity. 
The functions $f_i$ and $h_i$ describe the entity's evolution and measurement.
The discrete time axis is labeled by the index $k=0, 1, 2,\cdots$, in particular, with $k=0$ being regarded as 
the initial time. 
The vector $x = \mbox{col} (x_1,\cdots, x_N)$ is the lumped state of the whole system. 
The dynamics governing $x$ generally represent $N$ coupled entities as $f_i(x, u_i)$ depends on $x$. 
They also include  $N$ isolated entities as a special case 
when $f_i (x, u_i)$  depends on $x_i$ only, explicitly rewritten as $f_i (x_i, u_i)$.

The output signal $y_i(k), \; i=1,\cdots, N$, is quantized as $
y^{\rm q}_i(k) \inplus  \mathcal{Q}_{(n,m)}$, which is transformed into integer plaintext and hence 
encrypted before transmission over the network,
resulting in the ciphertext $\by_i(k)$. 
The postscript $\rm q$ represents quantization of the signal,
which rounds the elements of the signal to the nearest elements of $\mathcal{Q}_{(n,m)}$ less than or equal 
to the signal. 
A uniform quantization space $\mathcal{Q}_{(n,m)}$ is used for all entities $i=1,\cdots, N$, with the specified integers $n$
and $m$ depending on the storage capacity. 
This process is represented by the mapping \begin{align}
\by_i(k) = \cE_i \left( \left[\begin{array}{c} y^{\rm q}_i(k)\\
 -y^{\rm q}_i(k) \end{array}\right] \right)  
\label{en1}
\end{align}
where  $\cE_i$ contains the double-layer encryption, composite of $\cE_{R,i}$ and  $\cE_{P}$, as well as 
a rational-integer conversion $\cI_{n,m}$.
It is worth noting that both $y^{\rm q}_i(k)$ and $-y^{\rm q}_i(k)$ are encrypted when they 
are used in control design.   By doing so, any linear operation on 
$y^{\rm q}_i(k)$, $i=1,\cdots, N$, can be regarded 
as pure summation (not involving subtraction) on $y^{\rm q}_i(k)$ and $-y^{\rm q}_i(k)$, $i=1,\cdots, N$.

A control algorithm implemented in the remote controller is of the following form
\begin{align}
v_i(k) &= \kappa_i (z_i(k), \cD_1( \by_1(k)), \cdots, 
\cD_N(\by_N(k)) ) \nonumber \\
z_i(k+1) &= \varkappa_i (z_i(k), \cD_1( \by_1(k)), \cdots, 
\cD_N(\by_N(k)) )  \nonumber\\
z_i(0) &= \cE_o \left( \left[\begin{array}{c} \zeta_i(0)\\
-\zeta_i(0) \end{array}\right] \right)   
 \label{securecontrol}
\end{align}
where $v_i (k), z_i(k) \inplus \mathcal{Z}_{n_P^2}$,  $\zeta_i(0)\inplus \mathcal{Q}_{(n,m)}$ and the control functions $\kappa_i$ and $\varkappa_i$
rely on the received  ciphertext $\by_i(k)$ with a proper decryption, i.e.,  
$\cD_1( \by_1(k)), \cdots, 
\cD_N(\by_N(k))$. 
 The choice of $\zeta_i(0)$ is arbitrary. Note that $\cD_i$ decrypts the outer-layer encryption $\cE_{R,i}$, 
that is, 
\begin{align} \label{Diyi}
\cD_i( \by_i(k)) =  \cE_{o} \left( \left[\begin{array}{c} y^{\rm q}_i(k)\\
 -y^{\rm q}_i(k) \end{array}\right] \right). 
 \end{align}
Also, the functions  $\kappa_i$ and $\varkappa_i$ must be
properly designed such that their values are always within
the set $\mathcal{Z}_{n_P^2}$ to avoid a computation overflow; this is elaborated further later.

A general dynamic controller is used in the formulation, with $z_i(k)$ being the 
state of the dynamic compensator. 
In this paper, such a remote architecture is referred to as a networked control scheme. 
The  architecture is centralized if each controller unit relies on the ciphertexts from 
all the entities. It also contains decentralized control scenario 
when the functions $\kappa_i$ and $\varkappa_i$ rely on
the ciphertexts from some (not all) entities, according to a specified communication network topology. 

\begin{figure}[t]
 \centering
 \includegraphics[width=8cm]{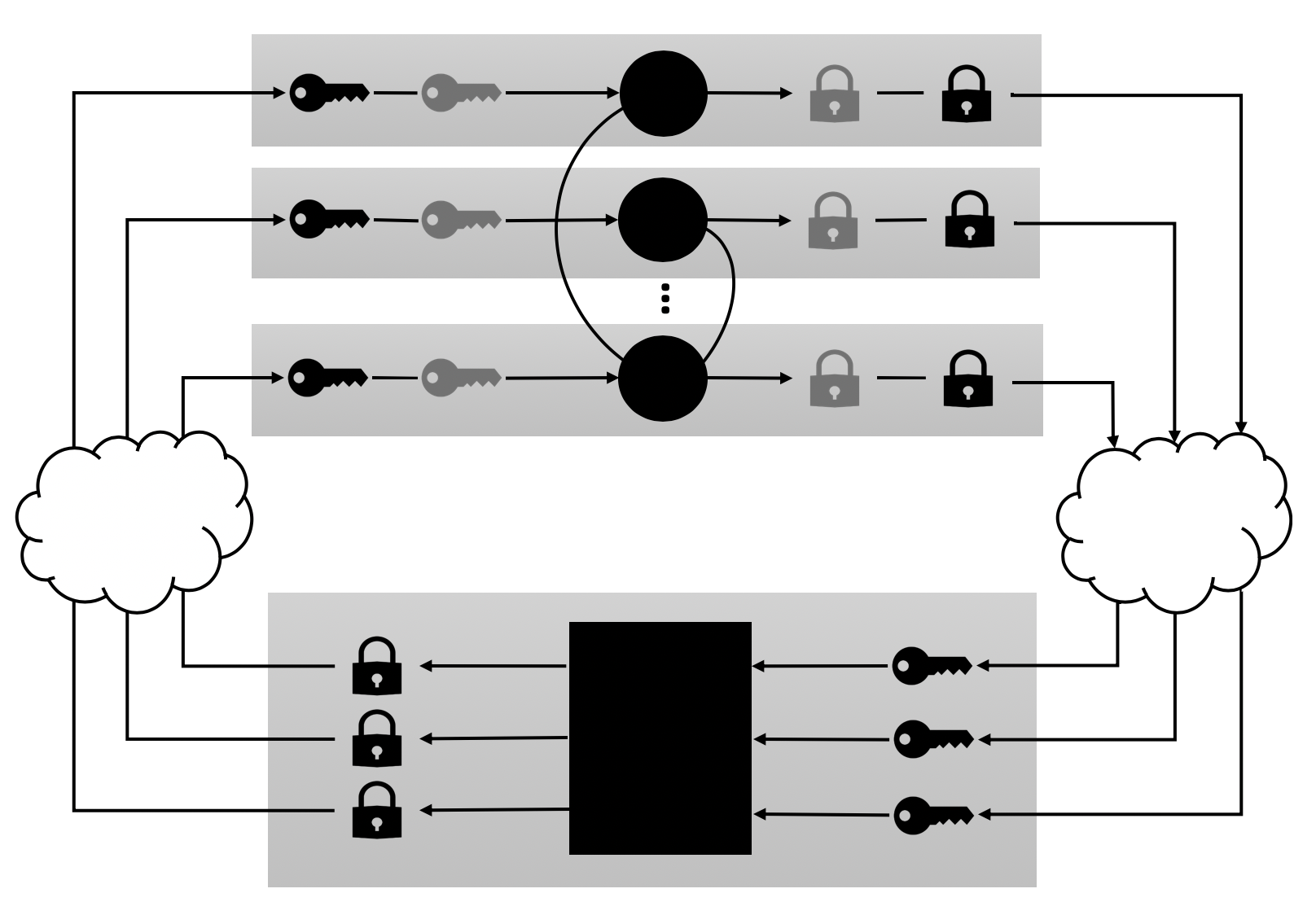} 
  \caption{Illustrative diagram of secure networked control for large-scale systems. The plant entities are represented 
by circles, the control units are within the remote rectangle, and the encryption and decryption protocols are represented 
by lockers and  keys, respectively.
  }
 \label{fig.diag}
\end{figure}

The designed control command $v_i(k)$ is again encrypted before transmission over network,
resulting in a ciphertext $\bv_i(k)$, represented by the mapping
\begin{align}\bv_i(k) = \bar \cE_i (v_i(k)).\label{en2}
\end{align}
Finally, the control command $u_i(k)$ is decrypted from $\bv_i(k)$ when it is received by the $i$-th entity, 
that is, 
\begin{align} u_i(k) = \bar \cD_i (\bv_i(k)),  \label{deco}
\end{align}
 where $\bar \cD_i$ contains a double-layer decryption composite of
$\bar \cE_{i}^{-1}$ and $\cE_{P}^{-1}$, as well as an
integer-rational conversion $\cI^{-1}_{n,m}$.

The networked control process for large-scale systems is explained above
and its architecture is also illustrated in Figure~\ref{fig.diag}.  The overall controller is composed of  (\ref{en1}), (\ref{securecontrol}), (\ref{en2}), and (\ref{deco}), which 
is called a private controller,  It, together with the plant (\ref{plant}), makes the complete closed-loop system denoted by
\begin{align}
 \bSigma =  (\ref{plant}) + (\ref{en1}) + (\ref{securecontrol}) + (\ref{en2}) + (\ref{deco}).
\end{align}
On the architecture of $\bSigma$, the main objective is stated below following the rigorous definitions of security
and some remarks.


\begin{Definition}
A signal $a(k)$ is {\it secure} in a ciphertext $\ba(k)$ to a party if the private decryption key is 
unavailable to the party.  
\end{Definition}

\begin{Definition} \label{def-security}
 The system $\bSigma$ is {\it secure} if the following two properties are satisfied:
 
 \begin{itemize}
 \item[(i)]  the signals $y^\q_i(k)$ and $u_i(k)$ are secure in the ciphertexts
$\by_i(k)$ and $\bv_i(k)$, respectively,  to all the entities and control units except the self-entity $i$;  

\item[(ii)]  the signals $y^\q_i(k)$ and $u_i(k)$
are secure in $\cD_i( \by_i(k))$ and $v_i(k)$, respectively, to all the control units.

\end{itemize}

 \end{Definition}

\begin{Remark} 
Encryption can protect the confidentiality of messages, which is the specific meaning of security in this paper. 
We call a message (signal) secure to an unauthorized party (adversary)
if the party can only access its ciphertext but not the decryption key.  
Therefore,  there exists no polynomial-time algorithm that an adversary can use to tell which of the messages could potentially be
the plaintext corresponding to a ciphertext more likely than the other to be the actual plaintext. 
On the architecture of $\bSigma$,  all the entities and control units are considered as curious-but-honest  adversaries. 
On one hand, there is mutual distrust between every components.
They aim to steal another entity's measurement output and control input, i.e., $y^\q_i(k)$ and $u_i(k)$. 
The information all the entities and control units can access includes the ciphertexts $\by_i(k)$ and $\bv_i(k)$.
The control units can also access the texts $\cD_i( \by_i(k))$ and $v_i(k)$.  This setting implies the 
two properties (i) and (ii) in Definition~\ref{def-security}. 
On the other hand, all the entities and control units are honest adversaries, that is, 
they run pass arbitrary messages without interfering with the outcomes.
Other techniques are still needed to protect the security in terms of 
integrity and authenticity of a message, which are however out of the scope of this paper.  
\end{Remark}

\begin{Remark} 
 The property (i) of Definition~\ref{def-security} considers the protection of information transferred over untrusted networks, that is, the protection of entity $i$'s information $y^\q_i(k)$ or $u_i(k)$ in the transmission ciphertexts
$\by_i(k)$ or $\bv_i(k)$. The information is protected from the controller and each of the entities within the system. 
The property (ii) addresses the protection of information between the system and the controller,
that is, the protection of entity $i$'s information $y^\q_i(k)$ or $u_i(k)$
during operations within the controller.
\end{Remark}

\begin{Remark}  The necessity of using a double-layer cryptographic strategy
has been mentioned in the motivating example at the beginning of Section~\ref{pre}. More specifically, 
this approach is used to ensure both properties (i) and (ii).  If a single-layer cryptographic mechanism is applied, 
there exist two typical scenarios, neither of which can achieve these properties.  (a)
The  decryption $\cD_i$ of the control unit is selected as the inverse function of $\cE_i$, that is, 
$\cD_i( \by_i(k))  = \cD_i(\cE_i(\mbox{col} (y^\q_i(k), -y^\q_i(k))) )   =\mbox{col} ( y^\q_i(k), -y^\q_i(k))$. Then the measurement $y^\q_i(k)$ can be directly 
used in the controller (\ref{securecontrol}), which much simplifies the controller design. 
However, in such a scenario,  a malicious control unit would easily detect the output measurement  $y^\q_i(k)$ 
of the entity.  So, (ii) is thus violated.
(b) In the second case, no decryption is applied at the control unit.  The ciphertext $\by_i(k)$ must use the same cryptosystem for all entities
to facilitate the design of an effective controller. As a result, 
(i) is violated, as 
the signal $y_i(k)$ does not remain secure with respect to other entities in the transmission ciphertext
$\by_i(k)$.
\end{Remark}

\medskip

{\it The objective of the paper is to find the encryption protocols
$\cE_P$, $\cE_{R,i}$, and $\bar \cE_{R,i}$,  and 
the  private control algorithm represented by the functions $\kappa_i$
and $\varkappa_i$, for $i=1,\cdots, N$,  such that
 
\begin{itemize}
\item{\bf P1:}  the  system $\bSigma$ has the desired control performance:
there exist $\zeta_i(k) \inplus \mathcal{Q}_{(n,m)},\; k\geq 0$
and a set ${\mathbb X}$, such that,
\begin{align*}
z_i(k) = \cE_o \left( \left[\begin{array}{c} \zeta_i(k)\\
-\zeta_i(k) \end{array}\right] \right)  
\end{align*}
and, with $x_c(k)=\mbox{col}(x(k), \zeta(k))$ and $\zeta = \mbox{col}(\zeta_1,\cdots, \zeta_N)$,
\begin{align}
\|x_c(k)\|\leq \alpha \rho^k \|x_c(0)\|+\varsigma 2^{-m} ,\; \forall x_c(0)\in \mathbb X, \label{stabilization}
\end{align}
for some $\alpha, \varsigma >0$ and $0<\rho<1$.

\item{\bf P2:}   the system $\bSigma$ is secure in the sense of Definition~\ref{def-security}.

\end{itemize}
}


\begin{Remark} 
The property P1 describes the desired control performance in two classes of problems, stabilization and 
consensus, which are rigorously formulated here and  will be explicitly investigated in Sections~V and VI, respectively. 
The formulation includes a residual error $\varsigma 2^{-m}$ as  the measurements are quantized for encryption, which will be 
further examined.  It is worth mentioning that the proposed encryption architecture is separated from the desired control performance, 
and hence can be applied to tackle security in more control problems and applications.
 The security property P2  has been explained in the remarks following Definition~\ref{def-security}.
\end{Remark}

\section{  A Double-Layer Cryptographic Architecture}\label{framework}

 A general secure networked control design paradigm using cryptographic mechanisms is proposed in this section.  In particular, the separation principle in  this paradigm enables the quantization-encryption-decryption protocol to be separated from the control algorithm design.
 This principle is stated in the following theorem.

\begin{Theorem}\label{thm-framework}
Consider the system $\bSigma$  satisfying the following three properties for two specified integers $n$ and $m$.

\begin{itemize}
\item (Controller subject to perturbation) There exists a controller
\begin{align} u_i(k) =&  \hat\kappa_i ( \zeta_i (k), y_1 (k) +\delta_1(k), \cdots,  \nonumber\\ &
y_N(k) + \delta_N(k) )   \nonumber \\
 \zeta_i(k+1) =& \hat\varkappa_i ( \zeta_i (k), y_1 (k) +\delta_1(k), \cdots, \nonumber\\ &
y_N(k) + \delta_N(k) )  \label{controller0}  \end{align}
 and a set ${\mathbb X}$, such that
the closed-loop system composed of (\ref{plant}) and (\ref{controller0}) has the desired control performance 
(\ref{stabilization})  for any $|\delta_i| <2^{-m}$. 
Moreover,  the signals $y_i(k), u_i(k), \zeta_i(k) \inplus [-2^{n-m-1},2^{n-m-1}), \; i=1,\cdots, N$.

 \item (Quantization property) 
The output signal $y_i(k)$ can be quantized as 
\begin{align} 
y^{\rm q}_i(k) \inplus  \mathcal{Q}_{(n,m)} ,\; 
 \|y_i(k) - y^{\rm q}_i(k)\|_\infty   <2^{-m},
\end{align}
and the functions $\hat\kappa_i $ and $\hat\varkappa_i$ satisfy 
 \begin{align}
\hat\kappa_i ( c_o, c_1, \cdots, 
c_N)  , \hat\varkappa_i (c_o, c_1, \cdots, 
c_N)   \inplus  \mathcal{Q}_{(n,m)}  \label{hatkappaQ}\end{align}
for $ \mbox{col}(c_o,c_1,\cdots, c_N )  \inplus
\mathcal{C}_{n,m}$ where the set  $\mathcal{C}_{n,m}$ is defined as follows
\begin{align*}
 \mathcal{C}_{n,m} :=& \{ \mbox{col}(c_o,c_1,\cdots, c_N ) \inplus \mathcal{Q}_{(n,m)} \:|\; \\
& \hat\kappa_i ( c_o, c_1, \cdots, 
c_N)  , \hat\varkappa_i (c_o, c_1, \cdots, 
c_N)   \inplus   \\&  [-2^{n-m-1},2^{n-m-1})
\}.
\end{align*}

\item (Cryptographic property) The functions $\kappa_i$, $\hat\kappa_i$, $\varkappa_i$, 
and $\hat \varkappa_i$, satisfy 
\begin{align}
& \cD_{o}( \kappa_i (\cE_{o}(c_o,-c_o), \cE_{o}(c_1,-c_1), \cdots, 
\cE_{o}(c_N,-c_N)) ) \nonumber\\ = & \hat\kappa_i (   c_o, c_1, \cdots, 
c_N)   \nonumber \\
& \varkappa_i (\cE_{o}(c_o,-c_o), \cE_{o}(c_1,-c_1), \cdots, 
\cE_{o}(c_N,-c_N))    \nonumber\\ =& \cE_o\left( \left[\begin{array}{c} \hat\varkappa_i (c_o, c_1, \cdots, 
c_N) \\
- \hat\varkappa_i (c_o, c_1, \cdots, 
c_N) \end{array}\right]\right) \label{homo}\end{align}
\end{itemize}
for $ \mbox{col}(c_o, c_1,\cdots, c_N )  \inplus
\mathcal{C}_{n,m}$. 
Then, the system $\bSigma$ 
achieves the property P1, when the initial condition satisfies
\begin{align} \mbox{col}(\zeta_i(0), y^\q_1(0), \cdots, 
y^\q_N(0)) \inplus \mathcal{C}_{n,m}. \label{ini}\end{align}
To simplify the notation, we use $\cE_{o}(c,-c)$  for $\cE_{o}(\mbox{col}(c,-c))$
throughout the paper. 
\end{Theorem}

\begin{Proof} Consider the closed-loop system  $\bSigma$  
composed of  (\ref{plant}), (\ref{en1}), (\ref{securecontrol}), (\ref{en2}), and (\ref{deco}). 
Let  $\delta_i (k) = y^{\rm q}_i(k)-y_i(k)$. From the quantization property, one has $\|\delta_i\|_\infty <2^{-m}$.

Assume there exists a constant $\bar k \geq 0$ such that 
\begin{align*}
z_i(k) = \cE_o \left( \left[\begin{array}{c} \zeta_i(k)\\
-\zeta_i(k) \end{array}\right] \right) 
\end{align*}
and
\begin{align}
\mbox{col}( \zeta_i(k), y^\q_1(k), \cdots, 
y^\q_N(k)) \inplus \mathcal{C}_{n,m},\; k=0,\cdots, \bar k. \label{bark}\end{align}
This statement is true with $k=0$; see the initial condition in (\ref{securecontrol}) and (\ref{ini}).
We will prove the statement for $k=\bar k+1$ and then apply mathematical induction. 

From \eqref{Diyi},  the first two equations in  (\ref{securecontrol}) become
\begin{align*}
v_i(k) = & \kappa_i (z_i(k),\cE_o (y^\q_1(k),-y^\q_1(k)), \cdots,  \\
& \cE_o (y^\q_N(k),-y^\q_N(k) ))  \nonumber \\
z_i(k+1)=& \varkappa_i  (z_i(k),\cE_o (y^\q_1(k),-y^\q_1(k)), \cdots,  \\
& \cE_o (y^\q_N(k),-y^\q_N(k) )). 
\end{align*}

The second equation further implies that, using the cryptographic property, and noting (\ref{bark}), we have
\begin{align*}
z_i(k+1) =& \varkappa_i (\cE_o(\zeta_i(k), -\zeta_i(k)),\cE_o (y^\q_1(k),-y^\q_1(k)),  \nonumber\\ & \cdots, 
 \cE_o (y^\q_N(k),-y^\q_N(k) ))\\
=& \cE_o\left( \left[\begin{array}{c}\hat\varkappa_i (\zeta_i(k), y^\q_1(k), \cdots, y^\q_N(k)) \\
- \hat\varkappa_i (\zeta_i(k), y^\q_1(k), \cdots, y^\q_N(k)) \end{array}\right]\right) \\ 
=& \cE_o \left( \left[\begin{array}{c} \zeta_i(k+1)\\
-\zeta_i(k+1) \end{array}\right] \right)
\end{align*}
for \begin{align}
\zeta_i(k+1) =  \hat\varkappa_i ( \zeta_i(k),y^\q_1(k), \cdots, y^\q_N(k) ). \label{prozeta}
\end{align}

From the cryptographic property again, one has the following equation, noting (\ref{bark}),
\begin{align*}
\cD_{o}(v_i(k)) = \hat\kappa_i ( \zeta_i(k),y^\q_1(k), \cdots, y^\q_N(k) )  \end{align*}
It is noted that $u_i(k) = \bar \cD_i (\bar \cE_i (v_i(k))) =\cD_{o}(v_i(k))$.
As a result,
\begin{align}
u_i(k) = \hat\kappa_i ( \zeta_i(k),y^\q_1(k), \cdots, y^\q_N(k) ). \label{prou}\end{align}
The set of equations (\ref{prozeta})-(\ref{prou}) is equivalent to  (\ref{controller0}), for $k=0,\cdots, \bar k$.
As a result, $y_i(\bar k+1)$ and $\zeta_i(\bar k+1)$ are determined by the closed-loop system 
composed of (\ref{plant}) and (\ref{controller0}).

From the quantization property, one has
$y_i(\bar k+1) \inplus [-2^{n-m-1},2^{n-m-1})$ and hence  $y^\q_i(\bar k+1)  \inplus \mathcal{Q}_{(n,m)}$, 
and also  $\zeta_i(\bar k+1)  \inplus \mathcal{Q}_{(n,m)}$.
Next, the property that $u(\bar k+1)$ and $\zeta_i(\bar k+1)$ for the system composed of (\ref{plant}) and (\ref{controller0})
are within the range $[-2^{n-m-1},2^{n-m-1})$ implies that
\begin{align*}
  \hat\kappa_i ( \zeta_i(\bar k+1),y^\q_1(\bar k+1), \cdots, y^\q_N(\bar k+1) )\\ \inplus [-2^{n-m-1},2^{n-m-1}) \nonumber \\
  \hat\varkappa_i ( \zeta_i(\bar k+1),y^\q_1(\bar k+1), \cdots, y^\q_N(\bar k+1)) \\ \inplus [-2^{n-m-1},2^{n-m-1}).
\end{align*}
As a result,
 \begin{align}\mbox{col}(  \zeta_i(\bar k+1), y^\q_1(\bar k+1), \cdots, 
y^\q_N(\bar k+1)) \inplus \mathcal{C}_{n,m}. \end{align}

By mathematical induction, one has (\ref{bark}) for all $k=0,1,2,\cdots$. Therefore, 
the closed-loop system under consideration is always equivalent to the one
composed of (\ref{plant}) and (\ref{controller0}). Hence, the desired control performance,
that is,  the property P1, is achieved.
\end{Proof}

\begin{Remark} \label{remark-flow} {   To guarantee the control performance in a secure control
environment, the prerequisite is the existence of the controller \eqref{controller0} with the same performance.
Additional conditions must also be satisfied to avoid a computation overflow or underflow when cryptographic mechanisms are applied. 
Firstly,  the signals $y_i(k), u_i(k), \zeta_i(k)$ must be within the range $[-2^{n-m-1},2^{n-m-1})$ 
such that it can be quantized with the error up to $2^{-m}$ in the space $\mathcal{Q}_{(n,m)}$ 
without overflow.
It is worth mentioning that reset of controller (e.g.,  the controller state is set to a publicly known number periodically
in \cite{murguia2020secure}) can be avoided in this paper.
Secondly, the functions $\hat\kappa_i$ and $\hat\varkappa_i$ must satisfy (\ref{hatkappaQ}), such that  
their values are rational numbers in base 2 without causing an underflow.
For this purpose, these functions take the linear form of integer coefficients, which are  explicitly constructed later. 
The third condition \eqref{homo} shows that the required functions $\hat\kappa_i$ and $\hat\varkappa_i$
can be equivalently realized by the functions $\kappa_i$ and $\varkappa_i$ applied on the ciphertexts, 
provided that the cryptosystem has a certain  homomorphic property and the aforementioned overflow or underflow issues 
are avoided.  Satisfaction of these conditions create additional challenges on the design of the controller, which will be explicitly analyzed 
in the subsequent sections.}
\end{Remark}

\medskip

Next, the specific encryption-decryption protocols are designed in the following theorem, which guarantees
 the cryptographic property in Theorem~\ref{thm-framework}, as well as the property P2.
 In the following,  ${\sum}$ and $\prod$ represent the normal summation and (element-wise) 
multiplication respectively. Also, for a real matrix $A=[a_{ij}] \in {\mathbb R}^{r_1 \times r_2}$ with $a_{ij}$ being the $(i,j)$-th entry,
and a vector $b=[b_1,\cdots b_{r_2}]^T$, let the following vectors be
\begin{align*}
 \prod  \cE_{o}^{|A|}(\sgn(A) b) & = 
 \left[ 
 \begin{array}{c}  \prod_{i=1}^{r_2}    \cE_{o}^{|a_{1 i}|}(\sgn(a_{1i}) b_i) \\
 \vdots\\
 \prod_{i=1}^{r_2}    \cE_{o}^{|a_{{r_1} i}|}(\sgn(a_{{r_1} i}) b_i)
 \end{array}
 \right],\\
  \prod  b^{|A|} &= 
 \left[ 
 \begin{array}{c}  \prod_{i=1}^{r_2}    b_i^{|a_1i|} \\
 \vdots\\
 \prod_{i=1}^{r_2}    b_i^{|a_{r_1i }|}
 \end{array}
 \right].
\end{align*}

\begin{Theorem} \label{thm-crypto} In Theorem~\ref{thm-framework}, suppose the control  
functions $\hat\kappa_i$ and $\hat\varkappa_i$ are linear with 
integer coefficients, that is, 
\begin{align}
\hat\kappa_i (c_o,c_1,\cdots, c_N) &=\sum_{j=0}^N \phi_{ij} c_j, \nonumber\\
\hat\varkappa_i (c_o,c_1,\cdots, c_N) &=\sum_{j=0}^N \varphi_{ij} c_j \label{linearkappa}
\end{align}
for integer matrices $\phi_{ij}$ and $\varphi_{ij}$.
If $\cE_P$, $\cE_{R,i}$, and $\bar \cE_{R,i}$  are designed as follows:

\begin{itemize}
{ 
\item  
 $\cE_P :  \mathcal{Z}_{n_P} \mapsto  \mathcal{Z}_{n_P^2}$ is a Paillier encryption with 
 \begin{align*}
 n_P > \max\{ 2^n \sum_{j=0}^N    |\phi_{ij}| \mathbf{1}  , \;  2^n \sum_{j=0}^N   |\varphi_{ij}|   \mathbf{1}  \}
 \end{align*} and the private key only available to all entities.


\item $\cE_{R,i} :  \mathcal{Z}_{n_R} \mapsto  \mathcal{Z}_{n_R}$ is an RSA encryption with $n_R \geq  n_P^2$ and the private key only available to the $i$-th controller.


\item $\bar \cE_{R,i}: \mathcal{Z}_{n_R} \mapsto \mathcal{Z}_{n_R}$ is an RSA encryption with the private key only available to the $i$-th entity.
}

\end{itemize}

\noindent Then, the cryptographic property in Theorem~\ref{thm-framework} is automatically satisfied for
\begin{align}
 \kappa_i (\cE_{o}(c_o,-c_o), \cE_{o}(c_1,-c_1), \cdots, 
\cE_{o}(c_N,-c_N))    \nonumber\\ = \prod_{j=0}^N    \cE_{o}^{|\phi_{ij}|}(\sgn(\phi_{ij}) c_j) \modd n_P^2,   \nonumber\\
 \varkappa_i (\cE_{o}(c_o,-c_o), \cE_{o}(c_1,-c_1), \cdots, 
\cE_{o}(c_N,-c_N))  \nonumber\\  =  \left[\begin{array}{c} \prod_{j=0}^N    \cE_{o}^{|\varphi_{ij}|}(\sgn(\varphi_{ij}) c_j) \modd n_P^2\\
 \prod_{j=0}^N   \cE_{o}^{|\varphi_{ij}|}(-\sgn(\varphi_{ij}) c_j)  \modd n_P^2  \end{array}\right] .  \label{homo2}\end{align}
Also the property P2 is achieved.
\end{Theorem}

\begin{Proof} We first prove the cryptographic property in  Theorem~\ref{thm-framework}.
One has
\begin{align*}
 &\cI_{n,m} (\hat\kappa_i (c_o, c_1, \cdots, 
c_N) )\\  =&2^m \sum_{j=0}^N \phi_{ij} c_j \modd 2^n  
\\ =&2^m \sum_{j=0}^N |\phi_{ij}| \sgn(\phi_{ij}) c_j \modd 2^n   
\\ =&  \sum_{j=0}^N |\phi_{ij}|  \cI_{n,m} (\sgn(\phi_{ij}) c_j) \modd 2^n .
\end{align*}
As $\cE_P$ is a Paillier encryption, one has 
\begin{align*}
 &\prod_{j=0}^N    \cE_{P}^{|\phi_{ij}|}(\cI_{n,m}(\sgn(\phi_{ij}) c_j))   \modd n_P^2  
 \\ =& \cE_{P}  ( \sum_{j=0}^N |\phi_{ij}| \cI_{n,m} (\sgn(\phi_{ij}) c_j)   )
\end{align*}
noting 
$0\leq \cI_{n,m}(\sgn(\phi_{ij}) c_j) < 2^n < n_P$
and
$0\leq \sum_{j=0}^N |\phi_{ij}|  \cI_{n,m} (\sgn(\phi_{ij}) c_j)  \leq
2^n \sum_{j=0}^N  |\phi_{ij}|   {\mathbf 1}< n_P$.
Then, 
\begin{align*}
& \cE_{P}^{-1} (\prod_{j=0}^N    \cE_{P}^{|\phi_{ij}|}(\cI_{n,m}(\sgn(\phi_{ij}) c_j)) \modd n_P^2)  \modd 2^n 
\\ =&  \sum_{j=0}^N |\phi_{ij}|  \cI_{n,m} (\sgn(\phi_{ij}) c_j)   \modd 2^n\\
= & \cI_{n,m} (\hat\kappa_i (c_o, c_1, \cdots, 
c_N) ) 
\end{align*}
and hence
\begin{align*}
 & \cI^{-1}_{n,m}(\cE_{P}^{-1}  (\prod_{j=0}^N    \cE_{o}^{|\phi_{ij}|}(\sgn(\phi_{ij}) c_j) \modd n_P^2) \modd 2^n ) \\ 
 =& \hat\kappa_i (   c_o, c_1, \cdots, 
c_N).
\end{align*}
As a result,
\begin{align*}
 \cD_{o}(\prod_{j=0}^N  \cE_{o}^{|\phi_{ij}|}(\sgn(\phi_{ij}) c_{j}) \modd n_P^2) =\hat\kappa_i (   c_o, c_1, \cdots, 
c_N).
\end{align*}
So, the first equation of (\ref{homo}) is proved noting (\ref{homo2}). 
The second equation of (\ref{homo}) can be proved using similar arguments. 
  
We can prove (i) of Definition~\ref{def-security} noting that, in 
$\by_i(k) =  \cE_{R,i}( \cE_{P} (\cI_{n,m}(y^{\rm q}_i(k),-y^{\rm q}_i(k)))$, 
the private key of $ \cE_{R,i}$ is unavailable to the entities and the private key of $\cE_{P}$ is unavailable to the control units;
and in $u_i(k) = \cD_{o} (\bar \cE_{i}^{-1}  (\bv_i(k)))$, 
the private key of $\bar \cE_{i}$ is unavailable to  all entities and control units except the self entity $i$.

Finally, we can prove  (ii) of Definition~\ref{def-security}  noting that, in $\cD_i (\by_i(k)) =  \cE_{P} (\cI_{n,m}(y^{\rm q}_i(k),-y^{\rm q}_i(k)))$, 
the private key of $\cE_{P}$ is unavailable to the control units; and 
in $u_i(k)  = \cI^{-1}_{n,m}(\cE_{P}^{-1} (v_i(k)) \modd 2^n )$, 
the private key of $\cE_{P}$ is unavailable to the control units.
\end{Proof}

\begin{Remark} {  
The importance of Theorem~\ref{thm-crypto} is two-fold. 
On the one hand, it describes the specific cryptosystems $\cE_P$, $\cE_{R,i}$, and $\bar \cE_{R,i}$, with their key generation rules;
on the other hand, it also gives the explicit construction of the controller functions $\kappa_i$ and $\varkappa_i$. 
It facilitates the practical implementation of the desired private controller. 
}
\end{Remark}

\begin{Remark} {  
The cryptographic techniques, especially the management of private keys in Theorem~\ref{thm-crypto} guarantee the property P2. 
If more than one entity is hacked and  becomes malicious, 
they cannot eavesdrop and obtain the information of the remaining benign entities, as they cannot break the outer-layer RSA encryption. 
If all the controllers are under attack and the RSA private keys are disclosed, the information of the entity $i$ within the system
still remains secure due to the inner-layer Paillier encryption. 
The secrecy of the entity $i$'s information is affected only when there is a malicious party who is able to obtain 
both the Paillier private key and the RSA private key.
}

\end{Remark}

\begin{Remark} Under Theorem~\ref{thm-crypto}, the controller (\ref{securecontrol}) with (\ref{en1}) takes
the following specific form
\begin{align}
v_i(k) = & \prod   (z_i^{\sgn(\phi_{i0})}(k))^{|\phi_{i0}|} \nonumber \\  & \times \prod_{j=1}^N   \cE_{o}^{|\phi_{ij}|}(\sgn(\phi_{ij}) y^\q_j(k))\modd n_P^2 \nonumber \\
 \left[\begin{array}{c}
  z_i^{+}(k+1) \\ 
    z_i^{-}(k+1)
  \end{array}\right]
 = &      \left[\begin{array}{c}   \prod  (z_i^{\sgn(\varphi_{i0})}(k))^{|\varphi_{i0}|}\prod_{j=1}^N   \cE_{o}^{|\varphi_{ij}|} \\
 (\sgn(\varphi_{ij}) y^\q_j(k)) \modd n_P^2\\
\prod  ( z_i^{-\sgn(\varphi_{i0})}(k))^{|\varphi_{i0}|}\prod_{j=1}^N   \cE_{o}^{|\varphi_{ij}|}\\ (-\sgn(\varphi_{ij}) y^\q_j(k)) \modd n_P^2  \end{array}\right]   \nonumber\\
 \left[\begin{array}{c}
  z_i^{+}(0) \\ 
    z_i^{-}(0)
  \end{array}\right] =& \cE_o \left( \left[\begin{array}{c} \zeta_i(0)\\
-\zeta_i(0) \end{array}\right] \right)
 \label{impcontroller}
\end{align}
where $z_i^{\sgn(\phi_{i0})} =z_i^+$ or $z_i^{-}$ for $\sgn(\phi_{i0})=+1$ or $-1$, respectively, 
in the element-wise manner.
When  $\phi_{i0} =0$,  $(z_i^{\sgn(\phi_{i0})}(k))^{|\phi_{i0}|}=1$ vanishes in the product. 
When $\sgn(\varphi_{i0}) =+1$ and $\sgn(\phi_{i0})=+1$, the controller reduces to 
\begin{align}
v_i(k) =&  \prod  (z_i^{+}(k))^{|\phi_{i0}|} \nonumber\\ & \times \prod_{j=1}^N   \cE_{o}^{|\phi_{ij}|}(\sgn(\phi_{ij}) y^\q_j(k)) \modd n_P^2\nonumber \\
  z_i^{+}(k+1)  
 =&\prod 
    (z_i^{+}(k))^{|\varphi_{i0}|} \nonumber\\ & \times \prod_{j=1}^N   \cE_{o}^{|\phi_{ij}|}(\sgn(\varphi_{ij}) y^\q_j(k)) \modd n_P^2   \nonumber\\
   z_i^{+}(0) =&  \cE_o (  \zeta_i(0) ),
\end{align}
with the $z_i^{-}(k+1)$-dynamics vanished. For the same reason, 
when $\sgn(\varphi_{i0}) =+1$ and $\sgn(\phi_{i0})=-1$, the controller reduces similarly with the $z_i^{+}(k+1)$-dynamics vanished. 
\end{Remark}

\section{Design of Controller}\label{case1}

Next, we provide a specific controller design for the secure stabilization problem for MIMO systems followed by an example.
The plant (\ref{plant}) takes the following linear form,  with the partition $u_i(k) =\mbox{col}(u_i^a(k), u_i^{b}(k))$ and 
$y_i(k) =\mbox{col}(y_i^a(k), y_i^{b}(k))$,
\begin{align}
x_i (k+1) =&  \sum_{j=1}^NA_{ij} x_j(k)  +B_i u^{\rm a}_i(k)  \nonumber \\
y_i^{\rm a}(k) =&   \gamma_1 C_i x_i(k) \nonumber \\
y_i^{\rm b}(k) =& \gamma_2 u^{\rm b}_i(k) , \; i=1,\cdots,N, \label{plant-stb}
\end{align} 
where all the matrices $A_{ij}$, $B_i$, and $C_i$ are rational matrices whose elements 
are rational numbers and $\gamma_1, \gamma_2>0$ are two parameters. 
The plant essentially consists of the upper two equations from the input $u_i^a$ to the output $y_i^a$.
The parameter $\gamma_1$ modifies the output to match the required resolution.
The third equation between $u^{\rm b}_i(k)$ and $y_i^{\rm b}(k)$
deliberately introduces an additional communication channel between the plant 
and the remote controller such that the state of remote controller can 
be sent back to the plant and processed 
by the gain $\gamma_2$, also to match the required resolution.   The so-called resolution 
has to be matched through $\gamma_1$ and $\gamma_2$, because 
only the integer gains can be implemented in the private controller design in 
the encrypted space, as elaborated in  Lemma~\ref{lemAc}.


 In this case, the proposed paradigm applies with $z_i(k) = z(k)$ and $\varkappa_i =\varkappa$, $i=1,\cdots, N$, and
the controller (\ref{securecontrol})  used 
in Theorem~\ref{thm-framework} takes the specific form  
with the linear functions  $\hat\kappa_i$ and $\hat\varkappa$  given as follows:
 \begin{align} 
  u_i^{\rm a}(k) =&    \sum_{j=1}^N \phi_{ij}  [ y^{\rm b}_j (k) +\delta^{\rm b}_j(k)]    \nonumber\\
   u_i^{\rm b}(k) =& (e_i \otimes I_i) \zeta(k) \nonumber\\
  \zeta(k+1) =&     \sum_{j=1}^N  \varphi_{oj}  [y_j^{\rm b} (k) +   \delta_j^{\rm b}(k) ]  +\sum_{j=1}^N \varphi_{j} [y^{\rm a}_j (k) +\delta^{\rm a}_j(k)]   \label{stabilizer}  
  \end{align}
for integer matrices $\varphi_o=[ \varphi_{o1},\cdots,  \varphi_{oN}]$, $\phi_{i}=[\phi_{i1},\cdots, \phi_{iN}]$,
$i=1,\dots,N$,  and $ \varphi=[\varphi_1,\dots,\varphi_N]$. 
  Here, $e_i = [0, \cdots, 0, 1, 0,\cdots, 0]$ whose $i$-th element is one
and $I_i$ is an identity matrix whose dimension is consistent with that of $x_i$.
It is noted that
$y_i^{\rm b} (k)  = \gamma_2  (e_i \otimes I_i) \zeta(k)$ and $\delta_i^{\rm b}(k)$  is the associated quantization error. Let $\delta_i (k) = \mbox{col}(\delta_i^{\rm a}(k) ,\delta_i^{\rm b}(k) )$. 
  It is worth mentioning that the controller coefficients represented by the 
matrices $\varphi_o$, $\phi_{i}$ and $\varphi$ must be integers to avoid 
a computation underflow as explained in Remark~\ref{remark-flow}.
The need for controllers having integer coefficients has also been discussed in literature such as in \cite{cheon2018} for PID controllers and FIR filters.

Representing the system and the controller in a compact form, let us denote $A=[A_{ij}]_{N\times N}$, $B=\mbox{block diag} (B_1,\dots,B_N)$,  $C=\mbox{block diag}(C_1,\dots,C_N)$,  $\phi=\mbox{col}(\phi_1,\dots,\phi_N)$, and define two matrices 
 \begin{align} A_c =\left[\begin{array}{cc} A & \gamma_2B  \phi \\ \gamma_1 \varphi C  & 
 \gamma_2\varphi_o \\ \end{array}\right] ,\;
 B_c=\left[\begin{array}{cc} 0 & B \phi \\ \varphi &  \varphi_o \\ \end{array}\right].\label{Ac}
 \end{align}

We first give a technical lemma under the following assumption.

\begin{Assumption}\label{Ass5}
The pairs $(A,B)$ and $(C, A)$ for the system~\eqref{plant-stb} are controllable and observable.
\end{Assumption}

\begin{Lemma}\label{lemAc} Under Assumption~\ref{Ass5}, 
there exist $\gamma_1, \gamma_2>0$ and integer matrices $\varphi_o$, $\phi$, and $\varphi$,
such that the matrix $A_c$ defined in (\ref{Ac}) is a Schur matrix.
\end{Lemma}

\begin{Proof} Under Assumption~\ref{Ass5},  there exist two matrices $K$ and $L$ such that 
 \begin{align*}
  \tilde A_c =& \left[\begin{array}{cc} A & BK \\ -LC & A+BK+LC\\ \end{array}\right] \\
 =& \left[\begin{array}{cc} I & 0 \\ I & I\\ \end{array}\right] 
  \left[\begin{array}{cc} A +BK & BK \\ 0 & A+LC\\ \end{array}\right]
  \left[\begin{array}{cc} I & 0 \\ -I & I\\ \end{array}\right] 
\end{align*}
 is a Schur matrix.
As rational numbers are dense, the  matrices $K$ and $L$  can be selected as rational matrices. 
Therefore, we can pick $\gamma_1, \gamma_2>0$ such that the following matrices 
\begin{align*} 
 \phi=K/\gamma_2,\; \varphi=-L/\gamma_1,\;  \varphi_o=(A+BK+LC)/\gamma_2
\end{align*}
are integer matrices.  Also, it is easy to check that $A_c$ and $\tilde A_c$ are identical, which are Schur matrices.
 \end{Proof}

Before we give the main result of this section, let us define some notations for the convenience of presentation. 
From Theorem 33 of Chapter 7d of \cite{callier2012linear}, for the Schur matrix $A_c$ defined in
(\ref{Ac}), there exist $0<\rho<1$ and $M>0$ such that $\|A_c^k\|\leq M\rho^k$. Let
 $\sigma = \|B_c\| d$ where $d$ is the dimension of $\delta(k)=\mbox{col}(\delta^{\rm a}(k),\delta^{\rm b}(k))$
 for  $\delta^{\rm a}(k) =\mbox{col}(\delta_1^{\rm a}(k),\dots,\delta_N^{\rm a}(k))$ and 
 $\delta^{\rm b}(k) =\mbox{col}(\delta_1^{\rm b}(k),\dots,\delta_N^{\rm b}(k))$.
  Let $x_c(k)=\mbox{col}(x(k), \zeta(k))$.

\begin{Theorem} \label{thm-MIMO}
Consider the closed-loop system composed of (\ref{plant-stb}) and (\ref{stabilizer}), under Assumption~\ref{Ass5}, 
with the parameters $\gamma_1, \gamma_2>0$ and integer matrices $\varphi_o$, $\phi$, and $\varphi$
selected in Lemma~\ref{lemAc}.  The closed-loop system achieves the desired control performance
\begin{align}
\|x_c(k)\|\leq M\rho^k \|x_c(0)\|+\frac{M\sigma}{1-\rho} 2^{-m} \label{xck}
\end{align}
for $|\delta(k)| < 2^{-m}$. Moreover, 
the signals $y_i(k), u_i(k), \zeta(k) \inplus [-2^{n-m-1},2^{n-m-1}), \; i=1,\cdots, N,$ 
if the initial condition satisfies $\|x_c(0)\| \leq R_o$ for 
\begin{align} 
R_o =& \frac{\beta}{M}-\frac{\sigma}{1-\rho} 2^{-m} ,\nonumber\\
\beta < &2^{n-m-1} \min\{1, \frac{1 -\|\phi_{i}\|  2^{-n+1} }{ \gamma_2 \|\phi_{i}\|}, \frac{1}{ \gamma_1 \|C_i\|}, 
  \frac{1}{\gamma_2} \}. \label{R0}
\end{align}
 \end{Theorem}

\begin{Proof}  Let $\bar y_i (k) = C_i x_i(k)= y_i^{\rm a}(k) /  \gamma_1$.
It is easy to check that the controller (\ref{stabilizer}) is equivalent to 
 \begin{align}  u_i^{\rm a}(k) =&  \gamma_2  \phi_{i}    \zeta(k) + \phi_{i}  \delta^{\rm b}(k)    \nonumber\\
  \zeta(k+1) =&     \gamma_2 \varphi_{o}    \zeta(k)  
  +\gamma_1  \sum_{j=1}^N \varphi_{j} \bar y_j (k)   \nonumber\\
&   + \varphi \delta^{\rm a}(k) + \varphi_{o}\delta^{\rm b}(k)  . \label{stabilizer2} 
   \end{align} 
Without considering the quantization errors,  the closed-loop system composed of
 \begin{align}
x_i (k+1) =& \sum_{j=1}^N A_{ij} x_j(k)  +B_i u^{\rm a}_i(k)  \nonumber \\
\bar y_i (k) =& C_i x_i(k) , \; i=1,\cdots,N. \label{plant-stb2}
   \end{align} 
and
 \begin{align} u_i^{\rm a}(k) =&  \gamma_2  \phi_{i}    \zeta(k)     \nonumber\\
  \zeta(k+1) =&     \gamma_2 \varphi_{o}    \zeta(k)  
  +\gamma_1  \sum_{j=1}^N \varphi_{j} \bar y_j (k)  
     \label{stabilizer3}    \end{align} 
can be proved to be stable. In fact, it can be put into the following compact form composed of
 \begin{align}
x(k+1) =& A x(k)+B u^{\rm a}(k) \nonumber \\
\bar y(k) =& C x(k) \label{plant-stbc}
   \end{align} 
and
 \begin{align}
u^{\rm a}(k) =& \gamma_2 \phi \zeta(k) \nonumber \\
\zeta(k+1) =& \gamma_2 \varphi_o \zeta (k)+\gamma_1 \varphi \bar y(k), \label{stabilizerc}
   \end{align} 
where $u^{\rm a}=\mbox{col}(u^{\rm a}_1,\dots,u^{\rm a}_N)$, $\bar y=\mbox{col}(\bar y_1,\dots,\bar y_N)$.
The controller \eqref{stabilizerc} is a typical Luenberger observer-based stabilizer. 
Furthermore,  the closed-loop system composed of \eqref{plant-stbc} and \eqref{stabilizerc} takes the following form
\begin{align}
x_c(k+1)=A_cx_c(k) \label{cl1}
\end{align}
for the Schur matrix $A_c$ given in (\ref{Ac}).  So, 
the closed-loop system~\eqref{cl1} is asymptotically stable, i.e., $\lim_{k\rightarrow \infty}x_c(k)=0$.

When the quantization errors are taken into consideration, 
the controller \eqref{stabilizer2} has  the following compact form
 \begin{align}
u^{\rm a}(k)=&\gamma_2 \phi\zeta(k)+ \phi \delta^{\rm b}(k)  \nonumber \\
\zeta(k+1)=& \gamma_2 \varphi_o \zeta(k)+\gamma_1 \varphi\bar y(k)+\varphi\delta^{\rm a}(k)+\varphi_o \delta^{\rm b}(k).   \label{stabilizerc2}
 \end{align}
 Then, the closed-loop system composed of \eqref{plant-stbc} and \eqref{stabilizerc2} takes the following form
\begin{align}
x_c(k+1) = A_cx_c(k)+B_c\delta(k) \label{cl2}
\end{align}
for $A_c$ and $B_c$ defined in (\ref{Ac}).
From \eqref{cl2}, we have 
\begin{align} 
x_c(k) = A_c^{k}x_c(0)+\sum_{j=0}^{k-1}A_c^{k-j-1}B_c\delta(j).
\end{align}
Noting $\|A_c^k\|\leq M\rho^k$ and $\sigma = \|B_c\|d$, the property (\ref{xck}) follows; see, e.g.,  \cite{jiang2001input}.
 
From  (\ref{xck}) and $\|x_c(0)\| \leq R_o$, one has 
  \begin{align*}
\|\zeta(k)\| \leq \|x_c(k)\|\leq M R_o +\frac{M\sigma}{1-\rho} 2^{-m} =\beta <2^{n-m-1}.
\end{align*}
More calculation shows
 \begin{align*} \|u_i^{\rm a}(k) \|_\infty  \leq &    \gamma_2 \|\phi_{i}\| \|\zeta(k)\|+\|\phi_{i}\|  2^{-m}   \nonumber\\
 \leq &  \gamma_2 \|\phi_{i}\| \beta+\|\phi_{i}\|  2^{-m}  <2^{n-m-1}
  \nonumber\\
  \| u_i^{\rm b}(k)\| =& \|\zeta(k)\|  <2^{n-m-1} \nonumber\\
\| y_i^{\rm a}(k) \| \leq&   \gamma_1 \|C_i\| \|x_i(k)\|\leq  \gamma_1 \|C_i\| \beta <2^{n-m-1} \nonumber \\
\|y_i^{\rm b}(k) \| =& \gamma_2 \|\zeta(k)\| \leq  \gamma_2\beta < 2^{n-m-1}.
 \end{align*}
It concludes that $y_i(k), u_i(k), \zeta(k) \in [-2^{n-m-1},2^{n-m-1})$, $i=1,\cdots, N$.
\end{Proof}

\begin{Remark}
In the desired performance  (\ref{xck}), the term $M\rho^k \|x_c(0)\|$ vanishes as $k\rightarrow \infty$,
so the closed-loop system state $x_c(k)$ is asymptotically bounded by $[{M\sigma}/({1-\rho}) ] 2^{-m}$ 
where $2^{-m}$ is sufficiently small for a sufficiently large $m$.  
The initial states can be sufficiently large for a sufficiently large $R_o$, and hence $\beta$, at the cost 
of a sufficiently large $n$.

\end{Remark}

\begin{Remark} Based on (\ref{stabilizer}), the implemented controller 
(\ref{impcontroller}) has the following specific form
 \begin{align}
v_i(k) =&  \left[ \begin{array}{c}  
  \prod_{j=1}^N  \cE_{o}^{|\phi_{ij}|}(\sgn(\phi_{ij}) [y_j^{\rm b}]^\q(k))\modd n_P^2 \\
(e_i \otimes I) z(k)
\end{array}\right]
  \nonumber \\
  z (k+1)  
 =& \prod_{j=1}^N \cE_{o}^{|\varphi_{oj}|}(\sgn(\varphi_{oj}) [y_j^{\rm b}]^\q(k))   \nonumber\\
&   \times \prod_{j=1}^N   \cE_{o}^{|\varphi_{j}|}(\sgn(\varphi_{j}) [y_j^{\rm a}]^\q(k)) \modd n_P^2  \nonumber\\
   z(0)  =& \cE_o \left( \zeta(0)  \right).
   \label{impstabilizer}
  \end{align}
\end{Remark}

By combining 
Theorems~\ref{thm-framework}, \ref{thm-crypto}, and \ref{thm-MIMO} and
noting that the quantization property is simply satisfied,  one has the following 
result.

\begin{Theorem} Consider the system (\ref{plant-stb}) with
the private controller  
 (\ref{en1}), (\ref{securecontrol}), (\ref{en2}), and (\ref{deco}), for two specified integers $n$ and $m$.
 Under Assumption~\ref{Ass5}, select the parameters $\gamma_1, \gamma_2>0$ and integer matrices $\varphi_o$, $\phi$ and $\varphi$ according to Lemma~\ref{lemAc} and hence 
 define the linear functions  $\hat\kappa_i$ and $\hat\varkappa$  as  
in (\ref{stabilizer}). Let $\cE_P$, $\cE_{R,i}$, and $\bar \cE_{R,i}$ be designed in Theorem~\ref{thm-crypto}.
Then, the controller (\ref{securecontrol}) with (\ref{en1}) takes the specific form (\ref{impstabilizer}).
For any initial condition satisfying $\|x_c(0)\| \leq R_o$ with $R_o$ given in (\ref{R0}), 
the closed-loop system achieves the desired control performance P1 in the sense of (\ref{xck}) and the security property 
P2. \end{Theorem}

\begin{figure*}[t]
 \centering
 \hspace*{-10mm}
 \begin{minipage}{0.48\textwidth}
 \includegraphics[width=9cm]{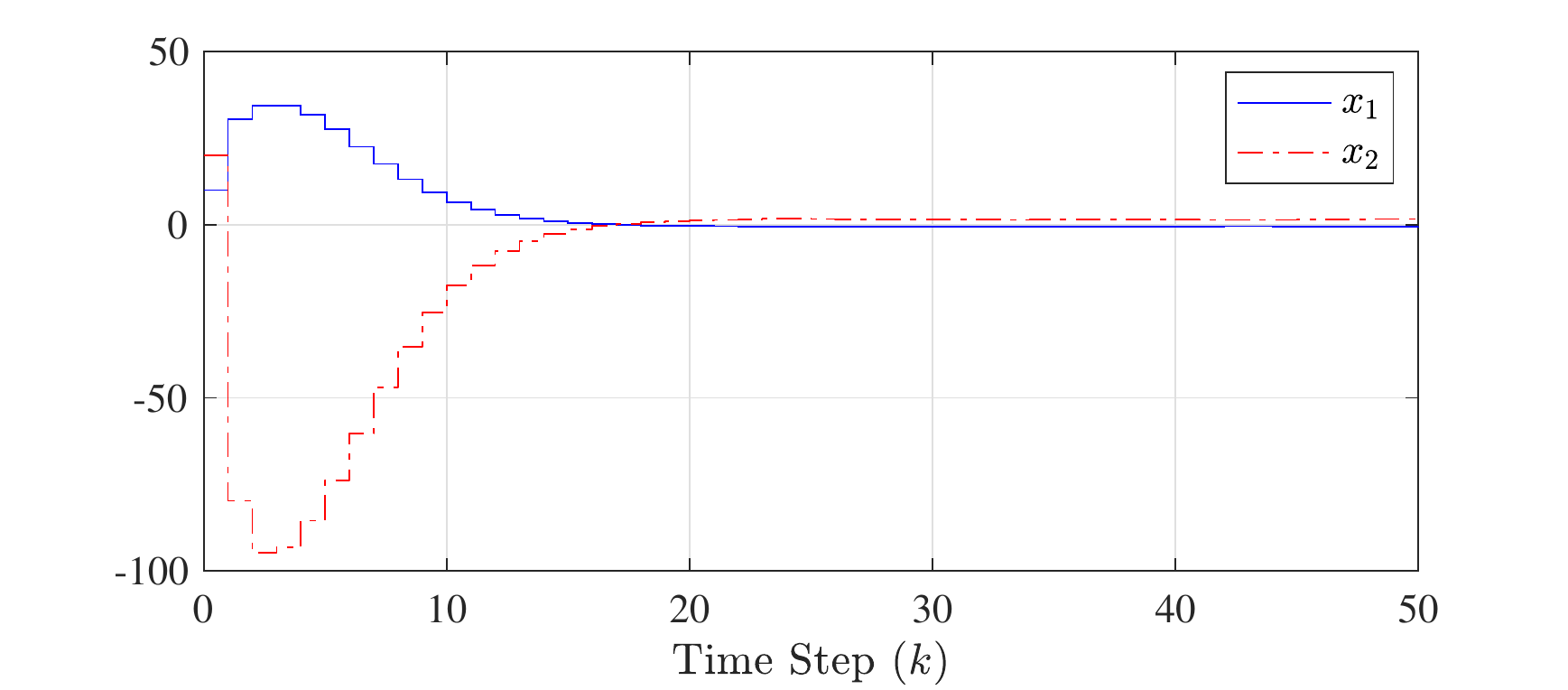} 
 \end{minipage}
 \hspace{2mm}
  \begin{minipage}{0.48\textwidth}
  \includegraphics[width=9cm]{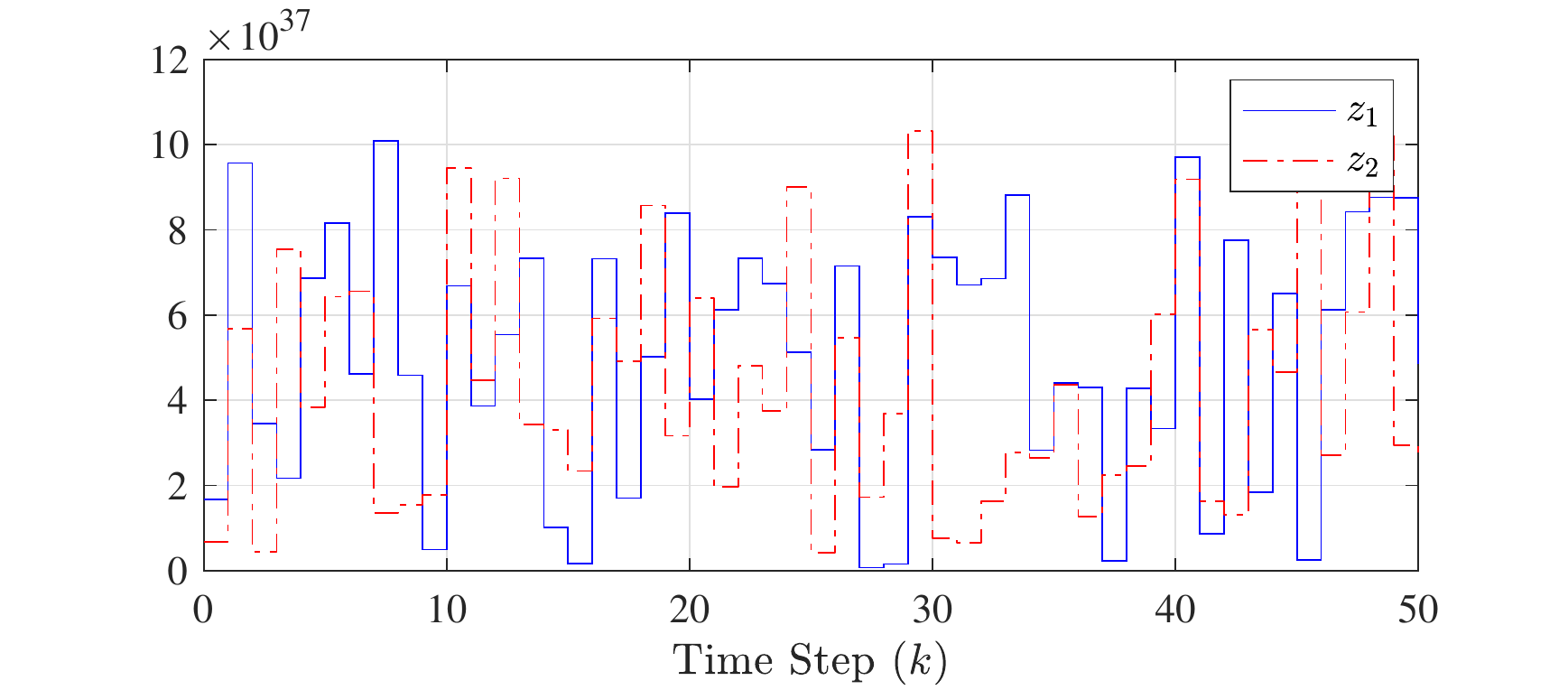} 
   \end{minipage}
   \caption{Left: profile of plant states practically approaching zero; right: profile of observer states in the encrypted space.}
   \label{xz}
\end{figure*}
\begin{figure*}[t]
 \centering
 \hspace*{-10mm}
 \begin{minipage}{0.48\textwidth}
 \includegraphics[width=9cm]{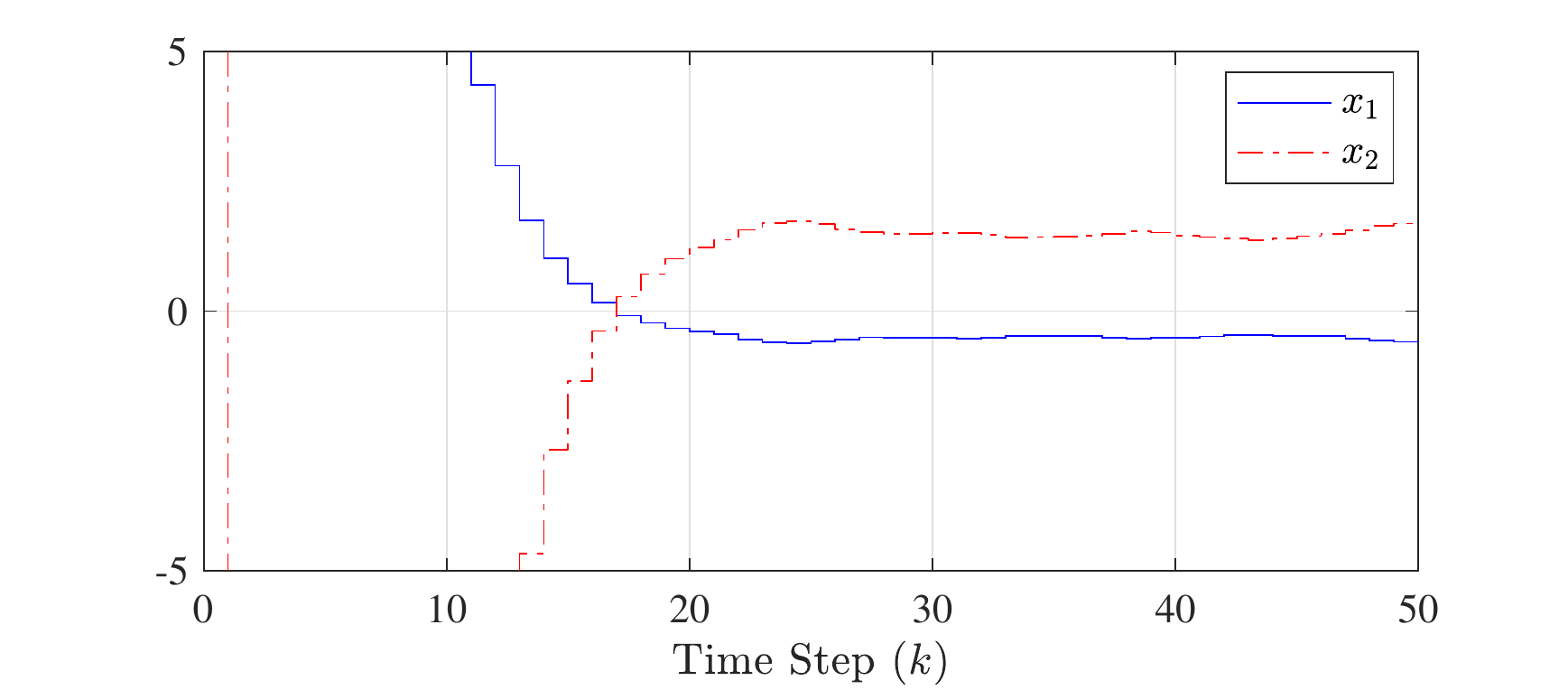} 
 \end{minipage}
 \hspace{2mm}
  \begin{minipage}{0.48\textwidth}
  \includegraphics[width=9cm]{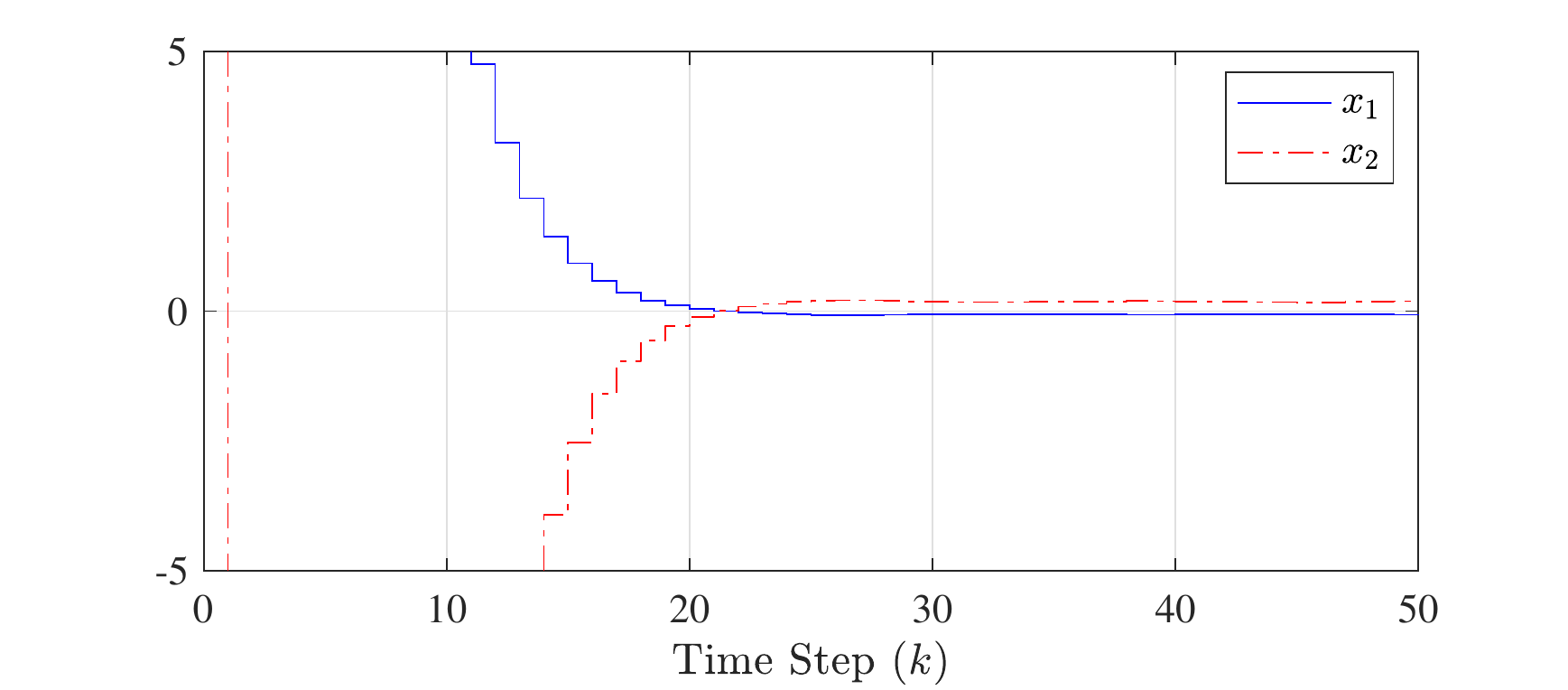} 
   \end{minipage}
  \centering\caption{Profile of residual errors of plant states approaching zero.
  Left: $m=6$; right: $m=9$.}
   \label{error}
\end{figure*}

\begin{Example} 
\label{example1}
 We consider the two-input two-output system (\ref{plant-stb}) of the following specific form
 \begin{align}
x_1(k+1)=&x_1(k)+x_2(k)+0.5u^{\rm a}_1(k) \nonumber \\
x_2(k+1)=&x_2(k)+u^{\rm a}_1(k)+u^{\rm a}_2(k) \nonumber \\
y^{\rm a}_1(k)=&\gamma_1 x_1(k)\nonumber \\
y^{\rm a}_2(k)=&\gamma_1 x_2(k).\label{MIMO-example} 
 \end{align}
It can also be put in the form (\ref{plant-stbc}) with $A=\left[\begin{array}{cc} 1&1\\0&1 \end{array}\right]$, $B=\left[\begin{array}{cc} 0.5&0\\1&1 \end{array}\right]$, and $C=\left[\begin{array}{cc} 1&0\\0&1 \end{array}\right]$. Obviously, Assumption~\ref{Ass5} holds, we can design the feedback gain and observer gain for the observer-based controller~\eqref{impstabilizer}, with the parameters $\gamma_1=0.1$, $\gamma_2=0.1$, and integer matrices $\phi=\left[\begin{array}{cc} 20&-10\\-65&-10 \end{array}\right]$, $\varphi=\left[\begin{array}{cc} -10&-12\\10&20 \end{array}\right]$, and $\varphi_o=\left[\begin{array}{cc} 30&17\\-55&-30 \end{array}\right]$.
The simulation results are illustrated in Fig.~\ref{xz}.
The left graph shows that the states of the system approach zero as expected by
the desired control performance. The observer is implemented in the encrypted space with 
the states plotted in the right graph. 
The presence of the observer states in disorder verifies that the privacy of the controller remain intact.
 The left graph of Fig.~\ref{error} zooms in on the trajectories of the system states and
shows the  residual errors when the states approach zero. The steady state errors are caused 
by quantization with $m=6$ and they can be reduced using a higher resolution.  For example, significantly smaller errors 
are shown in the right graph for a repeated simulation 
with $m=9$ and all the other conditions remain unchanged.

 The simulation was conducted 
with  $n=24$ and the initial conditions $x_1(0)=10$, $x_2(0)=20$, $\zeta_1(0)=12$, and $\zeta_2(0)=23$. 
The modular bits for generating keys of the first layer Paillier encryption are set to be $64$ and 
those for the second layer to be $256$. 
The simulation platform is MATLAB in a laptop computer of 2.2~GHz Intel Core i7 and 16~GB memory
integrated with an open-source C implementation of the Paillier cryptosystems \cite{libpaillier}.
The total simulation time including double-layer 
encryption/decryption (two encryption/decryption computations
in the outer layer and one encryption/decryption computation in the inner layer)
together with the control algorithm was recorded as $2.05$s for 51 instants, which implies that the average computation time 
for each control instant is $40.2$ms. It  can be well accommodated by one sampling 
period in many industrial applications such as control systems for chemical processes of typical sampling periods in seconds \cite{ZHANG2018616,WU2018185}.

\end{Example}

\section{Conclusion}\label{conclusion}
In this paper, we have established a secure networked control system design approach for large-scale cyberphysical systems using a novel double-layer cryptographic scheme. The approach is based on a design separation principle and is supported by a rigorous analytical proof.  In particular, a secure stabilization controller of MIMO systems has been presented with numerical simulations.   
The double-layer cryptographic scheme used Paillier and RSA cryptosystems to provide protection of information of 
plant entities and control units from any malicious party. Even though the approach is general, due to the specific use of Paillier, one can only handle the control laws involving linear computations. In the future work, we are planning to investigate other
homomorphic cryptosystems to cater for more complicated nonlinear control laws.  
 The computational performances of the Paillier and RSA algorithms are typically of the order of tens to hundreds of milliseconds \cite{jost2015,patil2016},   as also seen in the simulation example.
 This in turn implies that these computations in private controllers do not
cause any delay issue in the current discrete-time setting environment as they 
can be finished within one sampling period in 
many industrial applications such as chemical processes 
and smart grid networks. 
  Each cryptographic algorithm has its own advantages and it would be interesting to investigate different options for achieving greater efficiency in future research.

\bibliographystyle{ieeetr}

\bibliography{bibfile}

\end{document}